\documentclass[10 pt, conference]{IEEEtran}
\IEEEoverridecommandlockouts                       
\usepackage{url}
\usepackage[utf8]{inputenc}
\usepackage[T1]{fontenc}
\usepackage{amsmath,amssymb,amsfonts}
\usepackage{graphicx}
\usepackage{epstopdf}
\usepackage{subfigure}
\graphicspath{{\figures}}
\usepackage{textcomp}
\usepackage{xcolor}
\usepackage{verbatim}
\usepackage{makecell}
\usepackage{booktabs}
\usepackage{cite}
\usepackage{caption}
\usepackage{extarrows}
\usepackage{bm}
\usepackage{lettrine}
\usepackage[implicit=false]{hyperref}
\usepackage{amsthm}
\usepackage{algorithm,algorithmic}
\pdfoutput=1
\begin{document}
\title{
Base Station Beamforming Design for Near-field XL-IRS Beam Training} 

\author{
\large{Tao Wang},
\large{Changsheng You}, \IEEEmembership{\large{Member IEEE}}
and \large{Changchuan Yin}, 
\IEEEmembership{\large{Senior Member IEEE}}
\thanks{T. Wang and C. Yin are with the School of Information and Communication Engineering, Beijing University of Posts and Telecommunications~(BUPT), Beijing 100876, China (e-mail: taowang@bupt.edu.cn, ccyin@bupt.edu.cn).}
\thanks{C. You is with the Department of Electrical and Electronic Engineering, Southern University of Science and Technology~(SUSTech), Shenzhen 518005, China (e-mail: youcs@sustech.edu.cn).}
}
\vspace{0.5em}
\maketitle
\thispagestyle{empty}
\pagestyle{empty}

\begin{abstract}
Existing research on extremely large-scale intelligent reflecting surface (XL-IRS) beam training has assumed the far-field channel model for base station (BS)-IRS link. However, this approach may cause degraded beam training performance in practice due to the near-field channel model of the BS-IRS link. 
To address this issue, we propose two efficient schemes to optimize BS beamforming for improving the XL-IRS beam training performance. Specifically, the first scheme aims to maximize total received signal power on the XL-IRS, which generalizes the existing angle based BS beamforming design and can be resolved using the singular value decomposition (SVD) method.
The second scheme aims to maximize the $\ell_1$-norm of incident signals on the XL-IRS, which is shown to achieve the maximum received power at the user. To solve the non-convex $\ell_1$-norm maximization problem, we propose an eficient algorithm by using the alternating optimization (AO) technique.
Numerical results show that the proposed AO based BS beamforming design outperforms the SVD/angle based BS beamforming in terms of training accuracy and achievable received signal-to-noise ratio (SNR).
\end{abstract}

\begin{IEEEkeywords}
Intelligent reflecting surface, near-field communication, XL-IRS beam training.
\end{IEEEkeywords}

\section{Introduction}
Intelligent reflecting surfaces (IRS) have emerged as a promising technology to revolutionize the design of wireless networks\cite{c8, c4,IRS-Aided_Wireless_Communications_A_Tutorial}. 
Specifically, an XL-IRS consists of a large number of reflecting elements that can dynamically adjust signal reflection and/or amplitude. When deploying an XL-IRS, it is advantageous to place it near the base station (BS) due to several reasons\cite{co-site-IRS}. 
Firstly, placing the XL-IRS near the user equipment (UE) may incur high deployment costs as it requires densely deploying XL-IRSs to accommodate random UE mobility. 
Secondly, the channel between the BS and XL-IRS remains nearly static over shorter distances, which is favorable for channel estimation and information exchange between them. Hence, in this paper, we focus on the deployment of a BS-side XL-IRS, where the BS is in the near-field region of the XL-IRS\cite{liuYW2023NFC_Tutorial,you2023nearfield}.

In practice, XL-IRS beam training needs to be conducted to establish a high signal-to-noise ratio (SNR) BS-IRS-UE link. This training process involves two main steps: \textit{1)} the BS transmits pilot signals to the XL-RIS; and \textit{2)} the XL-IRS reflects the incident signals to sweep the beam towards the UE. 
Therefore, both the designs of BS beamforming and XL-XL-IRS beam training are crucial for the establishment of the BS-IRS-UE link.
Existing literature primarily focuses on codebook-based XL-IRS beam training schemes\cite{multi-beam-training-IRS, multi-IRS-beam-design, near-field-V1, codebook-IRS-Dai, Fast-Near-Field-Beam-Training-IRS, YouCS2023twostage_NFC}. For example, the authors in \cite{multi-beam-training-IRS, multi-IRS-beam-design} proposed a hierarchical XL-IRS codebook design to reduce the high overhead of far-field XL-IRS beam training. To address the near-field IRS-UE channel scenario, the authors in \cite{near-field-V1, codebook-IRS-Dai, Fast-Near-Field-Beam-Training-IRS, YouCS2023twostage_NFC} proposed various XL-IRS codebook designs and corresponding beam training strategies, aiming to improve training accuracy while reducing training overhead.
However, most existing works (e.g., \cite{YouCS2023twostage_NFC, multi-beam-training-IRS, near-field-V1, multi-IRS-beam-design, codebook-IRS-Dai}) assumed the simplified far-field BS-IRS channel model for XL-IRS beam training, where angle based BS beamforming is employed. This approach may cause degraded training performance due to the practical near-field channel model between the XL-IRS and BS.

To the best of our knowledge, this work presents the first attempt to address the design of BS beamforming during XL-IRS beam training.
Specifically, based on the spherical wavefront channel model, an optimization problem is formulated to maximize the received signal power at the UE by optimizing the BS beamforming. To address this problem, two efficient schemes are proposed, namely, the singular value decomposition (SVD) based BS beamforming and the alternating optimization (AO) based BS beamforming. The first scheme maximizes the total incident signal power on the XL-IRS using the SVD method. Next, by revisiting the formulated problem in the considered near-field channel model, we find that maximizing the UE received signal power is equivalent to maximizing the $\ell_1$-norm of the incident signals on the XL-IRS. 
Then, an efficient algorithm based on the AO technique is proposed to solve the reformulated non-convex problem.
Numerical results demonstrate that the proposed AO based BS beamforming design achieves higher training accuracy and improved received signal-to-noise ratio (SNR) compared to the SVD/angle based BS beamforming.

\textit{Notations}: scalar variables, column vectors and
matrices are denoted by normal-face letters, boldface lower letters and boldface upper-case letters, respectively. $[\cdot]_n$ denotes the $n$-th element of a vector. $[\cdot]_{n,:}$, $[\cdot]_{:,m}$ and $[\cdot]_{n,m}$ denote the $n$-th row vector, $m$-th column vector, and $(n,m)$-th element of a matrix, respectively. 
$(\cdot)^{\rm T}$, $(\cdot)^{\rm *}$, and $(\cdot)^{\rm H}$ denotes the transpose, complex conjugate, and conjugate transpose operators. $\angle (\cdot)$ normalizes the amplitude of each element in a matrix. $\mathbf{dist}(\cdot,\cdot)$ denotes the Euclidean distance between two coordinates. Finally, $\mathcal{CN}(\mu, \sigma^2)$ denotes the complex Gaussian distribution with mean $\mu$ and variance $\sigma^2$.
\vspace{-0em}
	\begin{figure}[t]
   \label{fig:system}
	\centering
	\includegraphics[scale = 1]{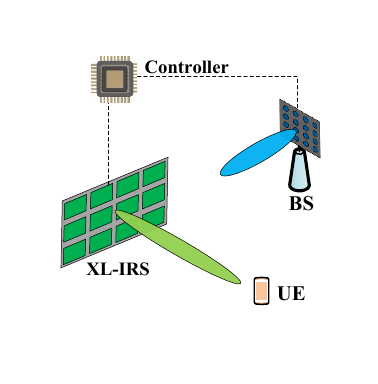}
	\vspace{-3em}
    \caption{An XL-IRS assisted wireless system.}
    \vspace{-0.5em}
	\end{figure}
\section{System Model}
We consider the downlink beam training for an XL-IRS assisted multiple-input single-output (MISO) system, as shown in Fig. 1. The direct BS-UE link is assumed blocked due to obstacle, whereas by properly deploying the XL-IRS, a BS-IRS-UE link is established. The system comprises a BS with $M$ antennas, denoted by the set $\mathcal{M} = \{0, 1, \cdots, M\}$. The XL-IRS consists of $N$ elements, denoted by the set $\mathcal{N} = \{0, 1, \cdots, N\}$. The UE is a single-antenna device located within the coverage area of the XL-IRS\footnote{The proposed BS beamforming design can be directly applied to XL-IRS beam training for multiple UEs.}. Furthermore, there is a controller attached to the XL-IRS, responsible for adjusting the reflection coefficients of the XL-IRS and facilitating information exchange with the BS through a separate reliable link. The coordinates of the $m$-th BS antenna, the $n$-th XL-IRS element, and the UE are denoted as $\boldsymbol{c}_m^{\rm{B}}, \boldsymbol{c}_n^{\rm{I}}$, and $\boldsymbol{c}^{\rm{U}}$, respectively. The distance between the $m$-th BS antenna and the $n$-th XL-IRS element is represented as $l_{n, m} = \mathbf{dist}(\boldsymbol{c}_m^{\rm{B}}, \boldsymbol{c}_n^{\rm{I}})$. Similarly, the distance between the $n$-th XL-IRS element and the UE is denoted as $d_{n} = \mathbf{dist}(\boldsymbol{c}^{\rm{U}}, \boldsymbol{c}_n^{\rm{I}})$.
\subsection{Signal Model}
We consider the propagation environment with limited scattering (typical for mmWave channels) and adopt the geometric free space propagation channel model to represent the BS-IRS channel, denoted as $\boldsymbol{H} \in \mathbb{C}^{N \times M}$. The mathematical formulation of the channel model is given by\cite{liuYW2023NFC_Tutorial}\footnote{The BS-IRS channel model used in this study is versatile enough to account for both near-field and far-field channels, accommodating a wide range of BS and XL-IRS array geometries.}

\begin{equation}   
 \label{BS-IRS_channel}
 \begin{split}
 [\boldsymbol{H}]_{n,m} &= \sqrt{\frac{G^{\rm B}
 A_{n,m}^{\rm I}}
 {4\pi l^2_{n,m}}}e^{j\frac{2\pi}{\lambda}l_{n,m}}, 
 \end{split}
 \end{equation}
where $G^{\rm B}$ represents the BS antenna gain, and $A_{n,m}^{\rm I}$ denotes the effective aperture of the $n$-th XL-IRS element with respect to the $m$-the BS antenna. Considering the fixed deployment of the BS and XL-IRS, the BS-IRS channel is assumed as quasi-static and can be independently estimated as a prior information\cite{Two-Timescale-Channel-Estimation-Dai}.
Similarly, the IRS-UE channel, denoted as $\boldsymbol{h}^{\rm H} \in \mathbb{C}^{1\times N}$, is represented as
\begin{equation}   
 \label{IRS-UE_channel}
 \begin{split}
 [\boldsymbol{h}^{\rm H}]_n &= \sqrt{\frac{G^{\rm I}
 A^{\rm U}}
 {4\pi d^2_{n}}}e^{j\frac{2\pi}{\lambda}d_{n}}, 
 \end{split}
 \end{equation}
where $G^{\rm I}$ denotes the XL-IRS element gain,
and $A^{\rm U}$ denotes the effective aperture of UE antenna. 
During XL-IRS beam training, the received signal of the UE is given by
\begin{equation}   
 \label{received_signal}
 \begin{split}
 y &= \boldsymbol{h}^{\rm H}
 \boldsymbol{\Phi}
 \boldsymbol{H}
 \boldsymbol{w}s+z,
 \end{split}
 \end{equation}
where $s$ denotes the unit-modulus transmitted symbol, $z \sim \mathcal{C}\mathcal{N}(0, {\sigma^2})$ denotes the received noise, $\boldsymbol{w} \in \mathbb{C}^{M\times 1}$ denotes the BS beamforming, $\bm{\Phi}={\rm{diag}}( e^{j\varphi_1}, \cdots, e^{j\varphi_{N}})\in \mathbb{C}^{N\times N}$ denotes the phase shifts introduced by the XL-IRS, where $\varphi_{n}$ denotes the phase shift of the $n$-th XL-IRS element. As such, the received power at the UE is given by
\begin{equation}
 \label{received-power}
 \begin{aligned}
  P &= \left|\boldsymbol{h}^{\rm H}
 \boldsymbol{\Phi}
 \boldsymbol{H}
 \boldsymbol{w}\right|^2+\sigma^2\\
 &=\left| \boldsymbol{v}^{\rm{T}}
 \boldsymbol{H}
 \boldsymbol{w}\right|^2+\sigma^2,
  \end{aligned}
\end{equation}
where $\boldsymbol{v}^{\rm{T}} \equiv \boldsymbol{h}^{\rm H}
\boldsymbol{\Phi}\in \mathbb{C}^{1\times N}$, and $[\boldsymbol{v}^{\rm{T}}]_n = \left|[\boldsymbol{h}^{\rm H}]_n\right|e^{j(\frac{2\pi}{\lambda}d_{n}+\varphi_n)}, n \in \mathcal{N}$.

\subsection{Existing BS Beamforming Design and Problem Formulation}
In conventional XL-IRS beam training procedure, the received power is measured to find the best $\boldsymbol{\Phi}$ within a predefined codebook $\mathcal{C}$, which can be formulated as
\begin{equation}   
 \label{IRS-training-procedure}
 \begin{split}
\max_{\boldsymbol{\Phi} \in \mathcal{C}}~& \left|\boldsymbol{v}^{\rm{T}}
 \boldsymbol{H}
 \boldsymbol{w}\right|^2+\sigma^2.
 \end{split}
\end{equation}
During the XL-IRS beam training, $\boldsymbol{h}^{\rm H}$, $\boldsymbol{H}$, and $\boldsymbol{w}$ are assumed to be fixed. Previous studies\cite{multi-beam-training-IRS,multi-IRS-beam-design,near-field-V1,codebook-IRS-Dai, Fast-Near-Field-Beam-Training-IRS, YouCS2023twostage_NFC} have primarily focused on angle based beamforming design for $\boldsymbol{w}$, assuming a single-path far-field channel model for $\boldsymbol{H}$, where the BS only needs to align its transmit beamforming with the angle of departure~(AoD) towards the XL-IRS. Consequently, the BS antenna array can be treated as a single antenna, leading to the omission of $\boldsymbol{w}$ optimization in this context.

However, for the considered XL-IRS aided communication systems with the XL-IRS deployed near the BS, it is crucial to accurately model the BS-IRS channel using a near-field MIMO channel model. This makes the existing angle based BS beamforming approach no longer applicable. Thus, this letter aims to investigate an efficient BS beamforming design for XL-IRS beam training.
To achieve this objective, we seek to maximize the received signal power at the UE by optimizing the BS beamforming configured during XL-IRS beam training, which can be formulated as
\begin{equation}   
 \label{BS-beamforming-problem-1}
 \begin{split}
\max_{\boldsymbol{w}}~&\left| \boldsymbol{v}^{\rm{T}}
 \boldsymbol{H}
 \boldsymbol{w}\right|^2,\\
 {\rm s.t.}~ &\Vert\boldsymbol{w}\Vert_2^2 \leq P^{\rm B},
 \end{split}
\end{equation}
where $P^{\rm B}$ denotes the maximum transmit power of the BS. It is evident that the objective function in (\ref{BS-beamforming-problem-1}) attains its maximum value when $\Vert\boldsymbol{w}\Vert_2^2 = P^{\rm B}$.
However, problem (\ref{BS-beamforming-problem-1}) is infeasible due to the unknown $\boldsymbol{v}^{\rm{T}}$ during XL-IRS beam training. In the next section, we propose two novel BS beamforming schemes for XL-IRS beam training.
\section{Proposed BS Beamforming Design for XL-IRS Beam Training}

\subsection{An Intuitive SVD based BS Beamforming}
During XL-IRS beam training, the accuracy of beam training can deteriorate when the received SNR is relatively low. Therefore, increasing the total incident signal power at the XL-IRS, represented by $\Vert
 \boldsymbol{H}
\boldsymbol{w}\Vert_2^2$, can enhance the performance of XL-IRS beam training. Based on this intuitive observation, the problem in (\ref{BS-beamforming-problem-1}) can be reduced as
\begin{equation}   
 \label{BS-beamforming-problem-2}
 \begin{split}
\max_{\boldsymbol{w}}~&\Vert
 \boldsymbol{H}
\boldsymbol{w}\Vert_2^2,\\
 {\rm s.t.}~ &\Vert\boldsymbol{w}\Vert_2^2 = P^{\rm B}.
 \end{split}
\end{equation}
This problem is equivalent to designing transmit beamforming in a conventional MIMO communication system with analog combining at the receiver. It can be readily solved using the SVD method; see the detailed procedures in Algorithm 1. It is worth noting that in far-field single-path BS-IRS channel conditions, where the rank of $\boldsymbol{H}$ is one, the only eigenvector of $\boldsymbol{H}$ is the same as the angle based BS beamforming vector. This implies that the existing angle based BS beamforming is a special case of the SVD based BS beamforming.

\floatname{algorithm}{Algorithm}
\begin{algorithm}[t]
	\caption{SVD based beamforming Design}
	\begin{algorithmic}[1]
 \STATE Input: $\boldsymbol{H}$.
  \STATE Output: $\boldsymbol{w}$.
  \STATE Performs SVD to $\boldsymbol{H}$ for the right singular matrix, $\boldsymbol{V}$
  \STATE Obtain the eigenvector corresponding to the largest singular value of $\boldsymbol{H}$, which is $[\boldsymbol{V}]_{:,1}$
  \STATE Return $\boldsymbol{w} = \sqrt{P^{\rm B}}[\boldsymbol{V}]_{:,1}$
  
	\end{algorithmic}
\end{algorithm}

\subsection{AO based BS Beamforming}
While the SVD based BS beamforming provides an intuitive solution to maximize the total incident signal power at the XL-IRS, it may not be optimal due to the neglect of $\boldsymbol{v}^{\rm{T}}$.
For the second proposed scheme, we reconsider problem (\ref{BS-beamforming-problem-1}), where $[\boldsymbol{v}^{\rm{T}}]_n = [\boldsymbol{h}^{\rm H}\boldsymbol{\Phi}]_n =  \left|[\boldsymbol{h}^{\rm H}]_n\right|e^{j(\frac{2\pi}{\lambda}d_{n}+\varphi_n)}, n \in \mathcal{N}$. We can assume that all elements of $\boldsymbol{h}^{\rm H}$ have the same amplitude, i.e. $\left|[\boldsymbol{h}^{\rm H}]_n\right| = a, n \in \mathcal{N}$. This assumption is reasonable for the line-of-sight (LoS) IRS-UE channel under both far-field and near-field scenarios\cite{Power-Scaling-Laws-and-Near-Field-Behaviors}. Further, let $\boldsymbol{\psi}^{\rm{T}} \equiv \angle \boldsymbol{v}^{\rm{T}}$, i.e. $[\boldsymbol{\psi}^{\rm{T}}]_n = e^{j(\frac{2\pi}{\lambda}d_{n}+\varphi_n)}, \forall~n\in \mathcal{N}$. Thus, we can write $\boldsymbol{v}^{\rm{T}} = a \boldsymbol{\psi}^{\rm{T}}$. Consequently, problem (\ref{BS-beamforming-problem-1}) is transformed into 
\begin{equation}   
 \label{BS-beamforming-problem-3}
 \begin{split}
\max_{\boldsymbol{w}, \boldsymbol{\psi}}~& a^2\left| \boldsymbol{\psi}^{\rm{T}}
 \boldsymbol{H}
 \boldsymbol{w}\right|^2,\\
 {\rm s.t.}~ &\Vert\boldsymbol{w}\Vert_2^2 = P^{\rm B}\\
 & \left|[\boldsymbol{\psi}^{\rm{T}}]_n\right| = 1, \forall~n\in \mathcal{N},
 \end{split}
\end{equation}
where $\boldsymbol{\psi}$ is regarded as an auxiliary constant modulus variable.
Problem (8) is a nonconvex optimization problem due to the constant modulus constraints on $\boldsymbol{\psi}^{\rm{T}}$ and $\boldsymbol{w}$. To address this, we propose an AO based algorithm that optimizes $\boldsymbol{\psi}^{\rm{T}}$ and $\boldsymbol{w}$ alternately in an iterative manner to obtain a local stationary solution. The steps of the proposed algorithm are outlined in Algorithm 2. The convergence of Algorithm 2 is proved in Proposition 1.
\floatname{algorithm}{Algorithm}
\begin{algorithm}[t]
	\caption{AO based beamforming Design}
	\begin{algorithmic}[1]
 \STATE Input: $\boldsymbol{H}$, a small $\epsilon$ denoting the convergence threshold.
  \STATE Output: $\boldsymbol{w}$.
  \STATE Randomly initialize $\boldsymbol{w}_{0}, {\boldsymbol{\psi}^{\rm{T}}_{0}}$, which satisfies $\Vert\boldsymbol{w}_{0}\Vert_2^2 = P^{\rm B}$ and $|[\boldsymbol{\psi}_{0}^{\rm{T}}]_{n}|  = 1, \forall~{n} \in \mathcal{N}$. 
  \STATE Initialize $f_{0} = \left| {\boldsymbol{\psi}^{\rm{T}}_{0}}
 \boldsymbol{H}
 \boldsymbol{w}_{0}\right|^2$.
        \STATE Set $i = 1$
    \WHILE{TRUE,}
		\STATE $\boldsymbol{\psi}^{\rm{T}}_{i} = \angle \left (\boldsymbol{H}
 \boldsymbol{w}_{i-1}\right )^*$
 \STATE $\boldsymbol{\widetilde{w}} = \left ({\boldsymbol{\psi}^{\rm{T}}_{i}} \boldsymbol{H}\right )^{\rm H}$
  \STATE $\boldsymbol{w}_{i} = 
\sqrt{P^{\rm B}} \boldsymbol{\widetilde{w}}/ \Vert \boldsymbol{\widetilde{w}}\Vert_2^2$
\STATE $f_{i} = \left| {\boldsymbol{\psi}^{\rm{T}}_{i}}
 \boldsymbol{H}
 \boldsymbol{w}_{i}\right|^2$
\IF{$|f_{i}-f_{i-1}|<\epsilon$}
  \STATE  \textbf{return }$\boldsymbol{w} = \boldsymbol{w}_{i}$
\STATE \textbf{break}
\ELSE
\STATE $i = i+1$
 \ENDIF
\ENDWHILE
	\end{algorithmic}
\end{algorithm}

\emph{Proposition 1:} Algorithm 2 convergences monotonically to a local stationary point.

\emph{Proof:} In Algorithm 2, the element-wise updates in step 7 and step 9 are optimal, which means that the cost function $\Vert {\boldsymbol{\psi}^{\rm{T}}_{i}}
 \boldsymbol{H}
 \boldsymbol{w}_{i}\Vert_2$~in step 10 is non-decreasing, i.e.,
\begin{equation}   
 \label{non-decreasing-proof}
 \begin{split}
\left| {\boldsymbol{\psi}^{\rm{T}}_{i+1}}
 \boldsymbol{H}
 \boldsymbol{w}_{i+1}\right|^2 > \left| {\boldsymbol{\psi}^{\rm{T}}_{i}}
 \boldsymbol{H}
 \boldsymbol{w}_{i}\right|^2, \forall~i.
 \end{split}
\end{equation}
This proves that Algorithm 2 is guaranteed to converge monotonically to, at least, a local stationary point. However, the convergence to the global optimal point cannot be guaranteed, due to the non-convexity of the original problem. $\hfill\blacksquare$

The problem (\ref{BS-beamforming-problem-3}) can be further understood in terms of its physical meaning. With a given $\boldsymbol{w}$, the maximum value of $\left| \boldsymbol{\psi}^{\rm{T}} \boldsymbol{H} \boldsymbol{w}\right|^2$ can be achieved as $\Vert \boldsymbol{H}\boldsymbol{w}\Vert_1^2$ when $\boldsymbol{\psi} = \angle (\boldsymbol{H} \boldsymbol{w})^{\rm{H}}$. This means that the problem (\ref{BS-beamforming-problem-3}) is equivalent to maximizing the $\ell_1$-norm of $\boldsymbol{H} \boldsymbol{w}$, leading to the following problem
\begin{equation}   
 \label{BS-beamforming-problem-4}
 \begin{split}
\max_{\boldsymbol{w}}~&\Vert\boldsymbol{H} \boldsymbol{w}\Vert_1,\\
 {\rm s.t.}~ &\Vert\boldsymbol{w}\Vert_2^2 = P^{\rm B}.
 \end{split}
\end{equation}
The $\ell_1$-norm maximization problem in (\ref{BS-beamforming-problem-4}) provides useful insights into the design of BS beamforming for XL-IRS beam training, which is elaborated in Remark 1.

\newtheorem{remark}{Remark}
\begin{remark}[Maximizing $\ell_1$-norm of incident signals on the XL-IRS]

\rm{
The optimization problem (\ref{BS-beamforming-problem-4}) implies that, in order to maximize the receiving power at the UE, the BS beamforming should aim to maximize the $\ell_1$-norm of the incident signals at the XL-IRS. This can be explained as follows.
The XL-IRS acts as a passive phased antenna array, applying analog beamforming to the incident signals transmitted by the BS. However, the XL-IRS cannot adjust the amplitudes of the incident signals. Therefore, there are two methods to enhance the received power at the UE served by the XL-IRS. 
First, increasing the total power of the incident signals on the XL-IRS, i.e., maximizing $\Vert\boldsymbol{H} \boldsymbol{w}\Vert_2^2$ as discussed in III-A. Second, evenly distributing the signal power across the elements of the XL-IRS to exploit the XL-IRS array gain. These two aspects are traded off by the BS beamforming design with the objective of maximizing $\Vert\boldsymbol{H} \boldsymbol{w}\Vert_1$, as illustrated in III-B.
}
\end{remark}

\subsection{Complexity Analysis}
For the angle based BS beamforming, the complexity of generating an angle based beamforming vector with a given angle information is $\mathcal{O}(N)$. For the SVD based BS beamforming, the complexity is determined by the SVD operation on $\boldsymbol{H}$, which is given by $\mathcal{O}(\min(MN^2, M^2N))$. For the AO based BS beamforming, the complexity is determined by the number of iterations, denoted as $I$, and the calculation of ${\boldsymbol{\psi}^{\rm{T}}_{i}}
 \boldsymbol{H}
 \boldsymbol{w}_{i}$ in each iteration, which is given by $\mathcal{O}(IMN)$. Note that for the BS-side XL-IRS deployment, the BS beamforming for XL-IRS beam training is assumed to be designed once for all, as the BS-IRS channel is assumed to be static\cite{co-site-IRS}.

\section{Numerical Results}

\begin{table}[t]
	\centering  
	\caption{Parameter configuration}  
	\label{parameters}  
	\begin{tabular}{cc|cc}  
		        \hline  
		& & & \\[-6pt]  
		Parameters&Values&Parameters&Values \\  
		        \hline
		& \\[-6pt]  
		$P^{\rm B}$&40 dBm&$\lambda$&0.01~m \\
		        \hline
    	& \\[-6pt]  
		$G^{\rm B}$ & 1 & $\sigma^2$ & -94~dBm \\
		        \hline
        & \\[-6pt]  
		$G^{\rm I}$&1&$A^{\rm U}$&$\lambda^2/(4\pi)$ \\
		        \hline
	\end{tabular}
\end{table}

In this section, we conduct Monte-Carlo simulations to validate the effectiveness of the proposed scheme. 
For the BS beamforming design, three schemes are compared: 1) \emph{Angle based beamforming}; 2) \emph{SVD based beamforming}; 2) \emph{AO based beamforming}. For the XL-IRS beam training, we adopt a two-phase beam training scheme as proposed in \cite{Fast-Near-Field-Beam-Training-IRS} to handle the near/far field IRS-UE channel.

In the simulation setup, we consider a parallel uniform linear array (ULA)\footnote{The ULA instead of uniform planar array (UPA) is used for simplicity. In addition, the proposed BS beamforming design is independent with the dimension of the BS-IRS channel. As such, the analytical results obtained for ULA can be readily extended to UPA structure.} for the BS antenna array and the XL-IRS\cite{liuYW2023NFC_Tutorial}, where $\boldsymbol{c}_m^{\rm{B}}, \boldsymbol{c}_n^{\rm{I}}$, and $A_{n,m}^{\rm I}$ can be obtained with basic geometry. 
The channel model described in Section II is adopted, with parameters listed in Table \ref{parameters}. Without specific indication, the BS-IRS distance is set as 5 m, the IRS-UE distance is set as 100 m, $M = 64$, and $N = 200$. 
The \textbf{anticipated SNR} is defined as the UE received SNR when the XL-IRS is configured with the perfect information of IRS-UE channel. 
The \textbf{training accuracy} is defined as the probability that the optimal codeword obtained from XL-IRS beam training is the best matching codeword with IRS-UE channel\cite{Fast-Near-Field-Beam-Training-IRS}. The \textbf{achievable SNR} is obtained from the XL-IRS beam training results. 


\begin{figure*}[t]
    \centering
    \subfigure[Convergence of Algorithm 2.]{\includegraphics[scale = 0.190]{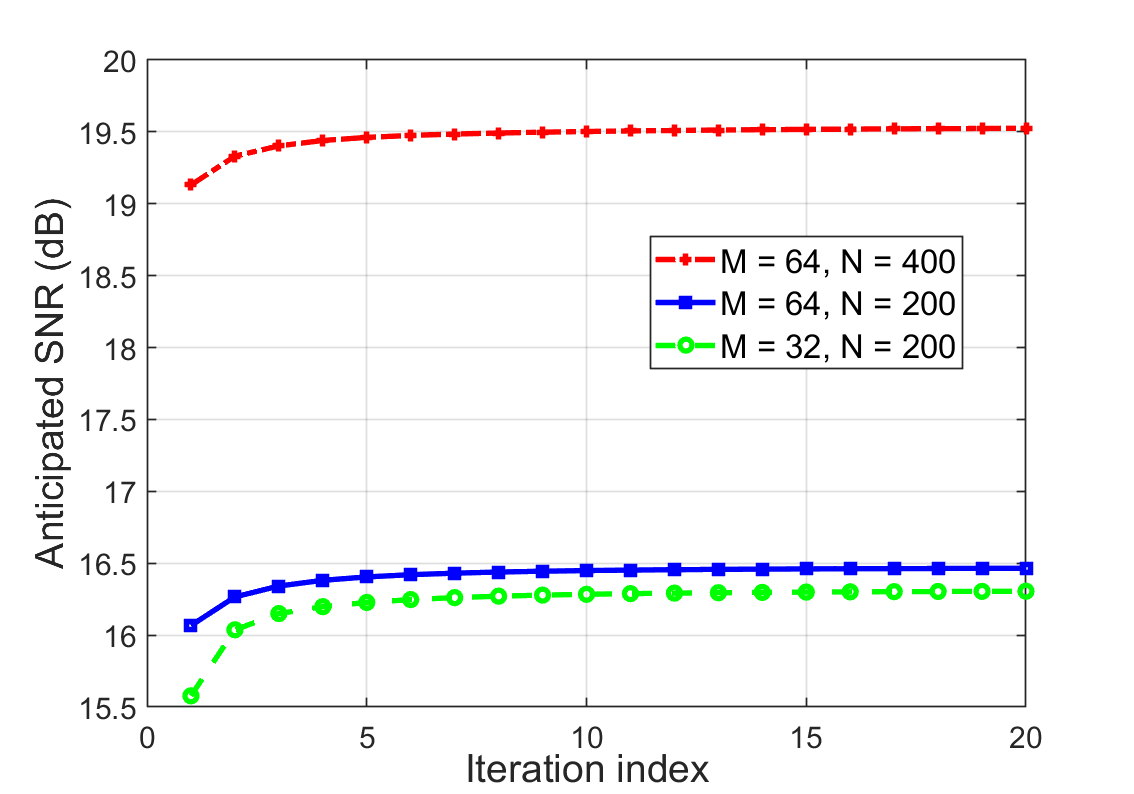}
	\label{fig:angle-SNR}}
    \subfigure[Power distribution across the XL-IRS.]
    {\includegraphics[scale = 0.190]{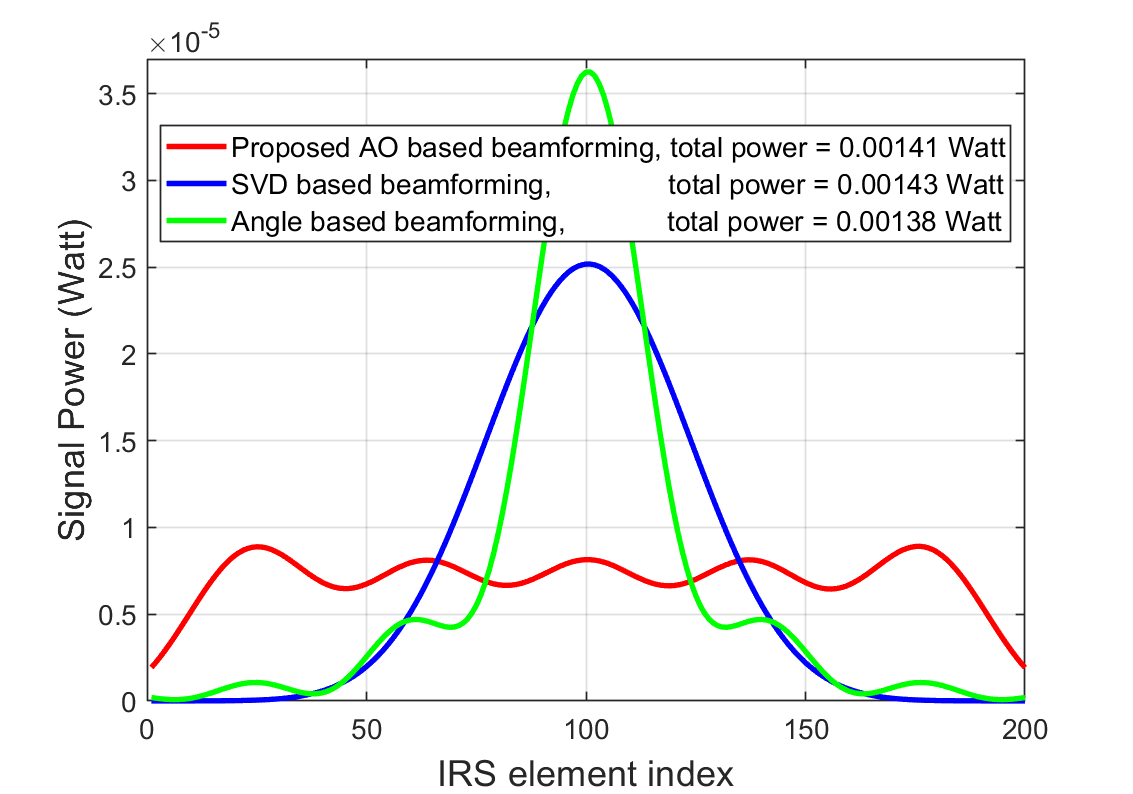}	
	\label{fig:distance-SNR}}
   \subfigure[Normalized XL-IRS DFT beam patterns under different BS beamforming schemes]
    {\includegraphics[scale = 0.285]{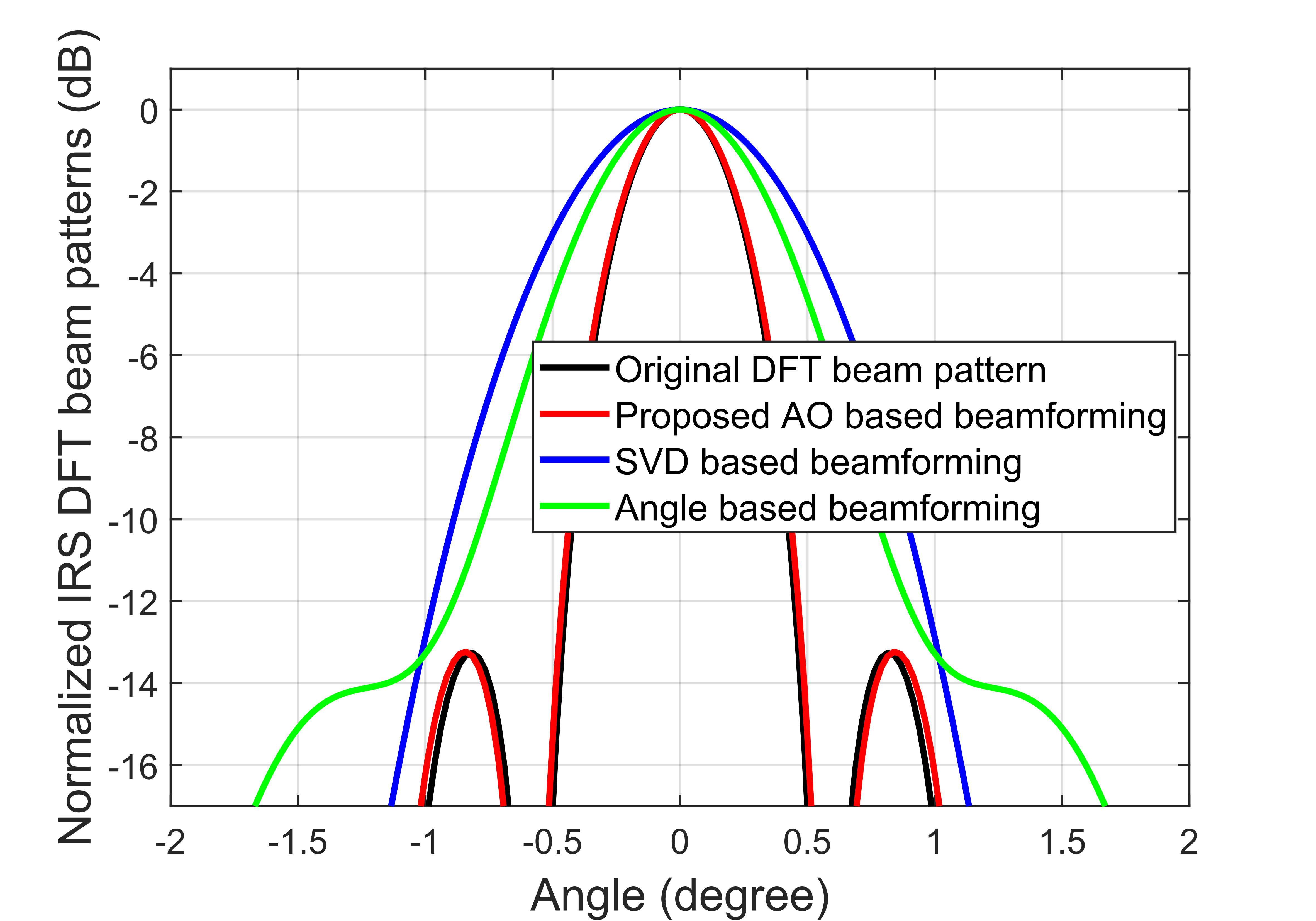}
\vspace{-0.5em}
\label{fig:IRS_beam_pattern}}
\quad    
\vspace{-0.5em}
    \centering
    \subfigure[Anticipated SNR v.s. BS-IRS distance.]
    {\includegraphics[scale = 0.285]{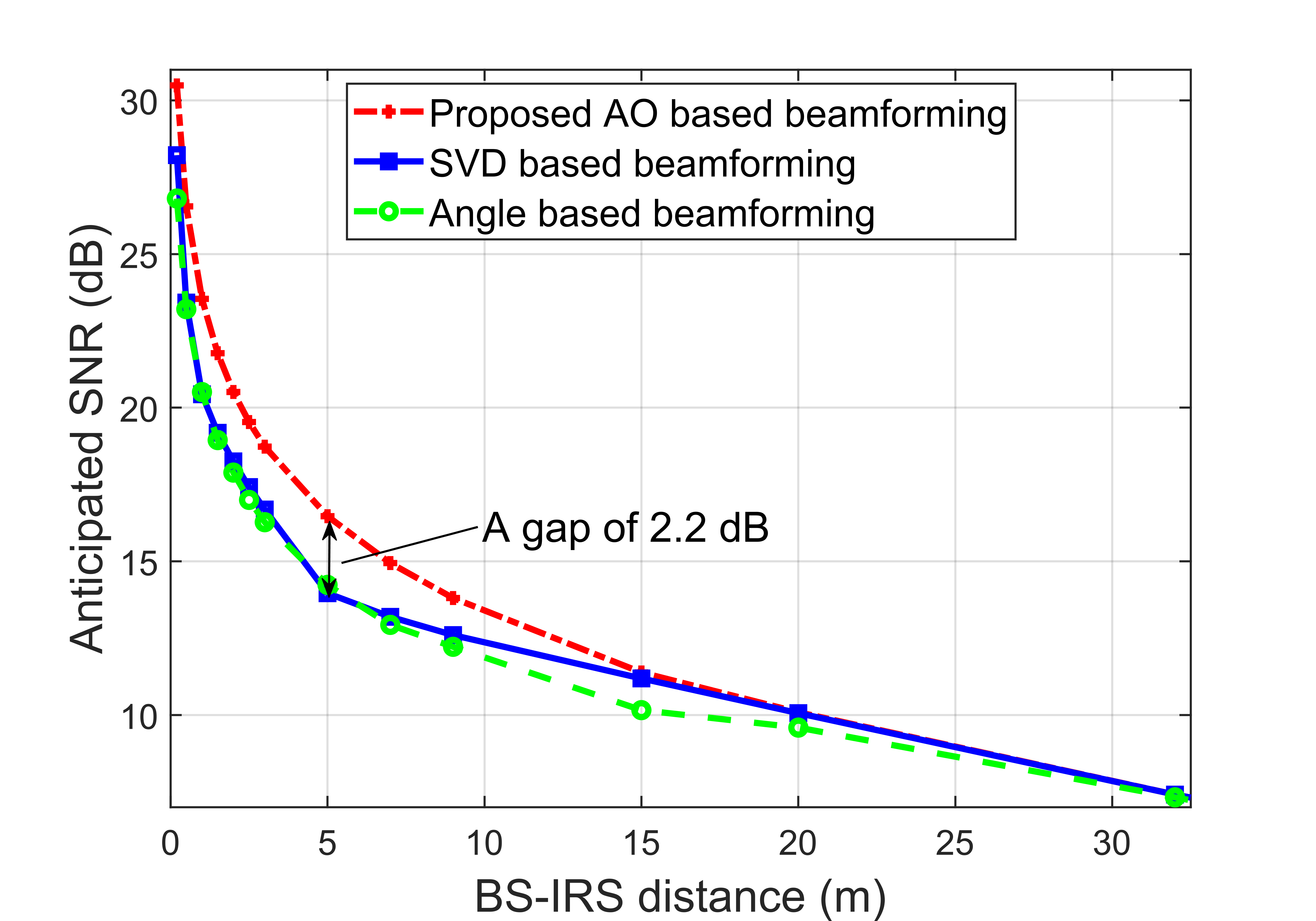}
    \label{fig:rate-SNR}}
    \subfigure[Training accuracy v.s. IRS-UE distance.]
    {\includegraphics[scale = 0.285]{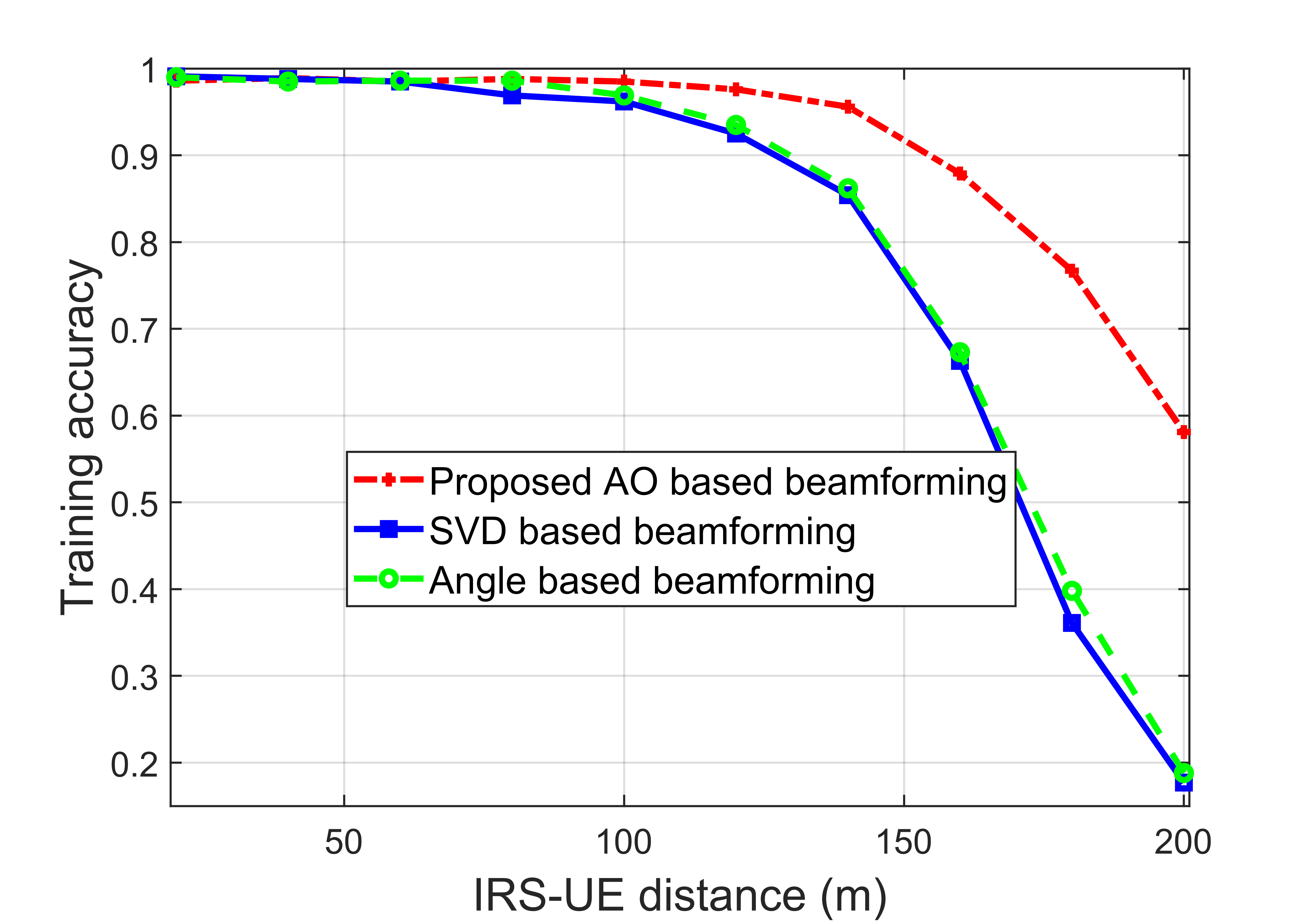}
	\label{fig:distance-distance}}
    \subfigure[Achievable SNR v.s. IRS-UE distance.]
    {\includegraphics[scale = 0.285]{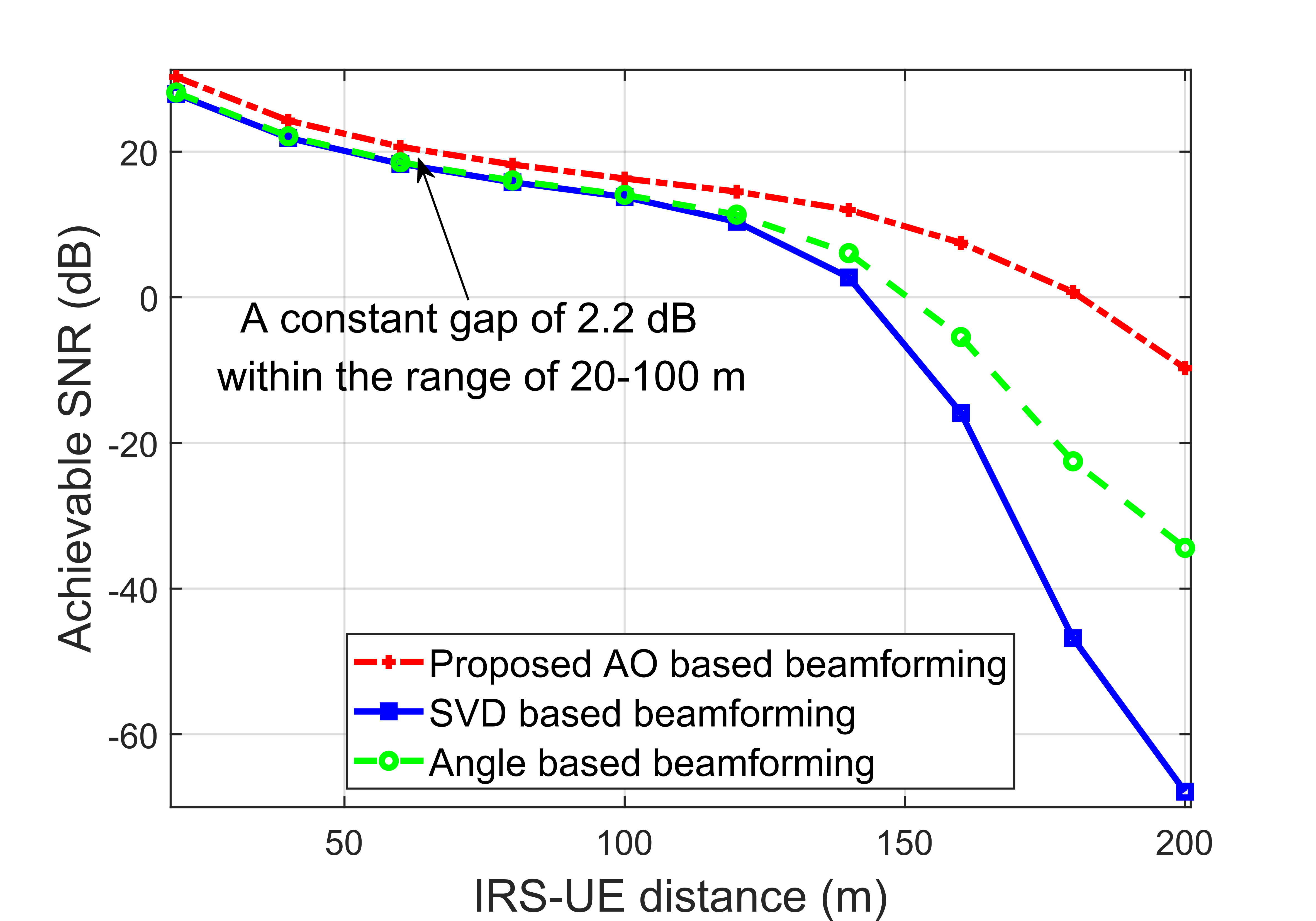}
	\label{fig:angle-distance}}

\caption{Convergence of Algorithm 2, and performance comparison between the proposed AO based beamforming, the SVD based beamforming, and the angle based beamforming.}
    \vspace{-0.5em}
    \label{simulation-results}
\end{figure*}

In Fig. 2(a), we observe that Algorithm 2 converges within 15 iterations for different values of $N$ and $M$. 
Fig. 2(b) shows the signal powers on the XL-IRS elements under different BS beamforming schemes. It is evident that compared to the SVD/angle based beamforming methods, the proposed AO based beamforming achieves a more even distribution of signal power across the XL-IRS elements. Besides, the total power on the XL-IRS remains similar among the beamforming schemes. This observation aligns with the analysis presented in Remark 1.
Fig. 2(c) shows the normalized XL-IRS discrete Fourier transform (DFT) beam patterns under different BS beamforming schemes. We notice that the XL-IRS beam pattern under the proposed AO based beamforming overlaps with the original DFT beam pattern. On the other hand, the XL-IRS beam patterns under the SVD/angle based beamforming methods get wider. This behavior can be explained by considering that the amplitudes of incident signals on XL-IRS are analogous to the weighting applied to XL-IRS beamforming vectors (codewords), which shapes the XL-IRS beam patterns\cite{amplitude_window}. The proposed AO based beamforming obtains an approximately equal-amplitude weighting (the red line in Fig. 2(c)), which has a minor impact on the XL-IRS DFT beam pattern. In contrast, the SVD/angle based beamforming methods produce bell-shaped amplitude weightings, leading to widened main lobes of the XL-IRS DFT beam patterns.

Fig. 2(d) demonstrates the impact of the BS-IRS distance on the anticipated SNR. It is observed that the anticipated SNR of all schemes decrease as BS-IRS distance increases, which is due to the reduced power captured by the XL-IRS. 
Moreover, the proposed AO based beamforming achieves a significant gain of up to 2.5 dB compared to the SVD based and angle based beamforming methods within the 0-30 m range. This result verifies that when the XL-IRS is located near the BS, the proposed AO based beamforming can significantly improve energy efficiency.

Additionally, further simulation results are presented in Figs. 2(e) and 2(f) to demonstrate the effects of the IRS-UE distance on training accuracy and achievable SNR of XL-IRS beam training. Specifically, the IRS-UE distance ranges from 20 m to 200 m, and the UE spatial angle is randomly distributed in [-1, 1] with 1000 beam training realizations. Several important observations can be made. First, the training accuracy and achievable SNR of all beamforming schemes decrease monotonically as UE distance increases, which is attributed to the more severe path loss. Second, the proposed AO based beamforming significantly outperforms the SVD based and angle based beamforming methods under various IRS-UE distances. Particularly, within the 20-100 m range, the gain achieved by the proposed AO based beamforming over the angle based beamforming is about 2.2 dB, which matches the anticipated gain in Fig. 2(d) when BS-IRS distance equals 5 m. This result validates the effectiveness of the proposed AO based BS beamforming for the XL-IRS beam training. Additionally, in the challenging long-distance (low-SNR) regime of 100-200 m, the proposed AO based beamforming exhibits more robust performance than other schemes in terms of both training accuracy and achievable SNR. This is because the SVD based and angle based beamforming methods result in bell-shaped amplitude weightings, which widen the main lobes of the XL-IRS DFT beam patterns\cite{amplitude_window} and consequently cause more severe performance degradation in the low-SNR regime.
\section{Conclusions}
\vspace{-0.5em}
In this letter, we studied the BS beamforming design for near-field XL-IRS beam training. 
Two efficient schemes were proposed, i.e. the SVD based BS beamforming and the AO based BS beamforming. The SVD based scheme maximizes the total incident signal power on the XL-IRS, while the AO based scheme seeks to maximize the $\ell_1$-norm of the incident signals on the XL-IRS. 
Numerical results demonstrated that the proposed AO based beamforming scheme outperforms the SVD-based and angle-based beamforming schemes in terms of training accuracy and achieved SNR. Furthermore, the proposed scheme is versatile and applicable to various BS-IRS array geometries and channel models.
\def\baselinestretch{0.85}
\bibliographystyle{IEEEbib}
\bibliography{IEEErefs}
\end{document}